# AUTOMATIC ORGANISATION, SEGMENTATION, AND FILTERING OF USER-GENERATED AUDIO CONTENT

*Gonçalo Mordido, João Magalhães, Sofia Cavaco*

NOVA LINCS, Departamento de Informática
Faculdade de Ciências e Tecnologia, Universidade Nova de Lisboa
2829-516 Caparica, Portugal
goncalomordido@gmail.com, {jm.magalhaes, scavaco}@fct.unl.pt

**ABSTRACT**

Using solely the information retrieved by audio fingerprinting techniques, we propose methods to treat a possibly large dataset of user-generated audio content, that (1) enable the grouping of several audio files that contain a common audio excerpt (*i.e.* are relative to the same event), and (2) give information about how those files are correlated in terms of time and quality inside each event. Furthermore, we use supervised learning to detect incorrect matches that may arise from the audio fingerprinting algorithm itself, whilst ensuring our model learns with previous predictions. All the presented methods were further validated by user-generated recordings of several different concerts manually crawled from YouTube.

## 1. INTRODUCTION

Given the abundance and ubiquity of video-oriented content (and, consequently, audio content) experienced in most social networks nowadays, it is important to understand such large amount of information in a meaningful way. One important step to achieve such understanding is to group the content in several clusters based on similarity, which in the context of this work is based on events. When we consider several user-generated recordings of different lengths reporting the same event, which is very likely to happen due to the nature of user-generated content, the existence of overlapping sections between two of such recordings means that they should belong to the same cluster/event.

Audio fingerprinting has been primarily used to detect if a given query song matches other songs in a preexisting database [1, 2, 6, 8]. Nonetheless, this algorithm retrieves very valuable information, that can be used for several other purposes. Here, we propose to use it to perform the organisation (clustering), segmentation and alignment, of audio recordings of music events.

The main contributions of this paper are then the organisation of a large dataset of audio recordings into the different events they portrait (section 2), with additional information regarding how each event's recordings are distributed over time (section 3), and the detection and filtering of incorrect matches possibly retrieved from the audio fingerprinting algorithm using a supervised learning approach (section 4), whilst ensuring our model learns with previous predictions (section 5).

Moreover, finding correlations between the content inside each cluster can also be very beneficial to achieve a better comprehension of the data. In this work we propose to align all event's song clips over time and we further use a quality inference technique already presented in previous work to order them in terms of their relative quality [7].

## 2. DATA ORGANISATION

Considering the abundant and ubiquitous nature of user-generated content, it is very likely to deal with a database of several different events (*e.g.* audio recordings of several concerts), in which each event has several recordings reporting it (*e.g.* relative to a certain concert song). Our goal is then to gather all song clips of a given event into the same cluster. Our technique to group clips of a given event is based on them having common excerpts of audio. Given the likely noisy and time sparse nature of user-generated content (*i.e.* the different recordings capture different parts of the event, with possible overlaps), we need to use a technique that is resistant to noise and at the same time can identify overlapping excerpts in music recordings from the same event.

Audio fingerprinting enables synchronising a query song $s_q$ against several other audio clips present in a formerly created database, whilst being relatively resistant to noise. Note that when we refer to a query song, we do not mean that we are dealing with the whole song, instead, we are referring to a portion of the whole song that has been recorded in an audio clip. Section 2.1 explains in more detail why using audio fingerprinting to characterise and compare the different audio files (similarly to what already proposed in our previous work [7]) is appropriate when dealing with possibly very noisy audio files, which is likely to be experienced in the context of our problem.

Once the audio fingerprints of the different audio files are compared and possibly matched, we use this information to identify overlapping excerpts and cluster our data into the different events. Section 2.2 takes a deeper look on how the grouping of the different recorded clips is indeed achieved and internally represented.

## 2.1. Audio Fingerprinting

The first step of our algorithm is to characterise the data with audio fingerprints. Using a fingerprint to characterise each recorded clip enables to efficiently represent and compare different clips, which is essential considering the vast occurrence of user-generated data experienced nowadays. Since the generation of this fingerprint involves the direct usage, or a combination of features from the audio signal, it is important to pick the features that are the most representative and, to some extent, invariant to distortion. Similarities between fingerprints of different song clips lead to a match of the clips.

Our algorithm uses Cotton and Ellis' landmark-based audio fingerprinting algorithm[1], which is based in the well-known approach formerly proposed by Wang [3, 8]. A fingerprint is composed of several landmarks, which in turn are generated through the analysis of two frequency peaks with high energy in a small period of time. More specifically, a landmark is a pair of two peaks, and contains information about each peak frequency, the time at which the first peak occurred, and the time offset between them.

Given a query song $s_q$, our algorithm uses the audio fingerprints information to match it against the song clips $s$ in the database [7]. Since each audio clip's fingerprints are a list of landmarks, our algorithm considers that two song clips, $s_q$ and $s$, contain the same audio excerpt if more than a certain number of landmarks are equal in both of their fingerprints; the threshold used is normally a small value (*e.g.* 5) since wrong matches are unlikely. To discover the time offset between the two clips we simply need to analyse the time difference between the timestamp of the equal landmarks in $s_q$ and $s$.

## 2.2. Audio Clustering

The second step of our algorithm is to organise the data into clusters, such that the clips from the same event (*i.e.* from the same whole music) are in the same cluster. This is achieved using the information retrieved from the fingerprinting stage. Moreover, since the different recordings will very likely be in different ranges of quality, from extremely noisy to clean

---

[1] This landmark-based audio fingerprinting algorithm is available in https://github.com/dpwe/audfprint.

---

recordings, audio fingerprinting permits the synchronisation of the low-quality recordings against better quality recordings in the database, conceivably in common audio portions that might not be too affected by noise in the low-quality recording.

In order to organise the audio clips in the database, we match each clip to all the other ones present in the database, ensuring all clips are tested against one another. In other words, for each clip in the database we consider it as a query song and use the fingerprints information to match it against all other clips in the database. Since there may be multiple clips for the same event (*i.e.*, whole song) each query song will likely have several song matches, that together compose the **matching list** of the query song.

The fingerprints information is used to build a graph, $G = (V, E)$, with the nodes, $V$, representing all song clips in our database and each edge in $E$ representing a match between two clips. Since each clip is represented by a node, we will use the same name ($s$) to refer to song clip $s$ and the node that represents that clip. Moreover, each edge is assigned a weight that consists of the offset (in seconds) between the two connected clips. In other words, if edge $(s_1, s_2, o_{12}) \in E$, then there is a match between clips $s_1$ and $s_2$ with offset $o_{12}$. Isolated nodes in the graph represent clips that have an empty matching list and that are not present in any of the other clips' matching lists. Even though the analysis of the weights of the paths is not necessary to performing the clustering, it will be essential to perform the audio segmentation presented in section 3.

The basis to detect and distribute the clips to the different clusters resides in the notion that if there is a path between two clips, then they should belong to the same cluster. This is an adaptation of Kennedy and Naaman's algorithm, that uses this graph-based representation to detect different episodes inside a given event [5].

## 3. AUDIO SEGMENTATION

Analysing how the different clips of each event are scattered over each event's timeline is of extreme importance to better manage the different audio files. Therefore this section focuses on finding the time intervals (*i.e.* segments) in which the different clips are distributed inside each cluster. An important aspect of this synchronisation task is that it only requires information already obtained from the audio fingerprinting algorithm that was used to perform the clustering described in section 2.

## 3.1. Audio Synchronisation

The offsets returned by the audio fingerprinting algorithm were further used to perform the alignment of the audio clips. This task uses the graph-based representation of our clips, $G$ described in section 2.2. As mentioned above, the weights of the edges are the offset (returned by the audio fingerprinting) between two clips. Following the paths in $G$, we can derive the offsets between any two clips in the same cluster by adding the weights in the path (*i.e.,* by calculating the cost of the path). It is important to notice that this only works because if there is a positive edge in the graph connecting two nodes, there is also a negative edge in the opposite direction. We can then represent the offset $o_{ij}$ between any two nodes $s_i$ and $s_j$ that are in the same cluster, as $o_{ij} = cost(G, s_i, s_j)$.

The actual way the synchronisation of all clips inside a cluster is made is by electing a **representative song clip** and by getting the offset of all the other clips relative to this one (that is, $o_{ir}$ for every song clip $s_i$ in the cluster). Note that the representative clip can be any of the cluster's clips, since all clips of a given event (cluster) are connected in the graph. After all offsets are obtained, if the representative clip is not the recording that has the earliest starting timestamp, the offset values are updated according to the clip with the earliest timestamp (*i.e.* the clip that starts first in the event's timeline).

We can define the earliest starting song clip $s_e$ as the clip with the minimum distance to the reference clip $s_r$:

$$\forall_{s_i \in V} \ o_{er} \leq o_{ir} .$$

This minimum distance can either be 0, if the representative song is indeed the earliest starting clip (since $o_{rr} = 0$), or a negative number, if $s_e$ starts before song $s_r$. Afterwards, we calculate all offsets $o_{ie}$. These can be obtained by adding the value of $o_{er}$ to the previously calculated offsets $o_{ir}$:

$$\forall_{s_i \in V} \ o_{ie} = o_{ir} + o_{er}$$

Using this approach, all offsets are greater or equal than 0 and correctly aligned in terms of their starting point along the event's timeline, since all offsets are now relative to the earliest starting clip.

## 3.2. Time-based Segmentation

By having the overall offsets of all clips of a given cluster, together with the duration of each clip, one has the knowledge of which clips exist in a given moment of time. Thus, we can organise an event with **segments**, such that segments coincide with the time interval of overlapping clips.

The overall event's timeline will be segmented into several non-overlapping segments. Given all offsets $o_{ie}$ (for all $s_i$ in the cluster), a new segment from time $t_{start}$ to time $t_{end}$ is created when one of the following situations occurs: (1) A new song clip $s_i$ starts at time $t_{start}$ ($o_{ie} = t_{start}$). (2) A clip $s_i$ with duration $d(s_i)$ ends at time $t_{end}$ (that is, $o_{ie} + d(s_i) = t_{end}$). As a consequence, whenever a new segment starts at $t_{start}$, there is a segment ending at $t_{start-1}$, except when $t_{start} = 0$ meaning that it is the first segment of that event.

The song clips can then be cut according to the timestamps of each of the segments they are part of. For instance, if song $s_1$ belongs to segment $A$ and $B$, then the song is cut into song $s_{1A}$ and $s_{1B}$ ($s_1$ is equal to the concatenation of $s_{1A}$ and $s_{1B}$).

This information is encapsulated in a tuple that represents a segment. The tuple contains an initial and final timestamp, and all clips that overlap between that period of time, $(t_{start}, t_{end}, s_{1A}, s_{2A}, \ldots)$. Each cluster, or event, is then composed by several segments, that give information on which clips are available in the different time intervals and therefore at any moment of time in the event's timeline.

## 3.3. Quality Inference

In previous work we proposed a method to infer the quality of each song clip relative to all the other clips inside a given cluster by analysing the sum of each clip's number of matching landmarks against the rest of the clips in the database [7]. This method can be further used to infer the quality of the clips inside each segment by matching them using the audio fingerprinting algorithm (that is, the algorithm is called once more but with the clips within the segment and not all clips in the database).

However, given the possible small time length of the segments, together with the possible small number of clips within each segment, matches are less likely to happen. Thus, increasing the number of landmarks by increasing the number of landmarks/sec performed by the algorithm for each clip, generates a higher number of matching landmarks between different song clips and therefore increases the likeliness of matches to occur.

Since the clusters were formed based on song clips with common excerpts, and after the filtering of false matches that will be presented in section 4, we can eliminate the matching landmarks threshold leading a match to be declared even with only 1 matching landmark between two clips. Since all clips inside a segment are time-aligned, the expected offset returned by the algorithm should be 0 seconds, meaning all the other matching landmarks with different offsets can be discarded and not considered for the clip's quality score.

This quality inference step enables ultimately for song clips to be ordered based on their relative quality inside each segment. Thus, on top of having information to which clips are available at a given time in the overall event's timeline, we now know how the different song clips inside the segment relate in terms of their relative quality.

## 4. FILTERING METHOD

Even though unlikely, the probability of a false match between two clips from the audio fingerprinting algorithm is still greater than 0. We propose a method to filter out such false matches from the clusters.

In previous work, we proposed a filtering approach based on the analysis of significant drops on the derivatives of the percentage of matching landmarks between the query and matched song relative to the overall number of the matched clip's landmarks [7]. Here, we present an alternative method that uses machine learning to detect such false matches.

### 4.1. Feature Selection

Our samples, or feature vectors, are derived from the fingerprinting algorithm's output, and every song is represented by several samples. Each sample corresponds to a match returned by the fingerprinting algorithm.

Given a query song $s_q$, the fingerprinting algorithm returns the following for every song $s_i$ in the database: (1) the number of landmarks, $\#L_{s_q}$, of the query song $s_q$, (2) the offset between $s_q$ and $s_i$, that is $o_{qi}$, (3) the number of matching landmarks with offset $o_{qi}$, which we call $\#ML_{o_{qi}}$, and (4) the number of total matching landmarks in all offsets, $\#TML$. Note that when a song is added to the database, the number of landmarks computed for that song is also retrieved from the algorithm, hence the number of landmarks of all songs are known. Thus, (5) the number of landmarks, $\#L_{s_i}$, of song $s_i$ is also known. Since the actual value of the offset does not directly influence if a match is correct or incorrect, it is not considered to enter the feature space. However, all the other referred features might be a good indicator of a false match.

The set of available and possibly relevant features, for each pair $(s_q, s_i)$, is then the following:
$F = \{\#ML_{o_{qi}}, \#TML, \#L_{s_q}, \#L_{s_i}\}$. We tested our models with several subsets of $F$, more specifically:

- $\{\#ML_{o_{qi}}, \#TML\}$
- $\{\#ML_{o_{qi}}, \#TML, \#L_{s_q}\}$
- $\{\#ML_{o_{qi}}, \#L_{s_q}, \#L_{s_i}\}$
- $\{\#ML_{o_{qi}}, \#TML, \#L_{s_q}, \#L_{s_i}\}$

Each one of our classifiers was trained with these features subsets to access which combination generates the best model.

### 4.2. Training Data

Since the goal of our models is to predict whether a sample is a false match or a true match, there are only two classes: 0 and 1, respectively. **False matches** are incorrect matches. These can be **wrong matches**, if the two matched songs do not have any common audio excerpt, or **repetition matches**, if they have indeed a common excerpt but the assigned offset is not correct. The latter case can be easily detected as it happens when a song $s_i$ appears in the matching list of a query $s_q$ several times with different offsets, described as repetitions. In this case, the match offset ($o_{qi}$) with the highest number of matching landmarks is considered a true match (*i.e.,* assigned to class 1), whilst all other match offsets ($o'_{qi}, o''_{qi}, o'''_{qi}, \ldots$) are considered false matches (*i.e.,* assigned to class 0).

A dataset of 198 audio recording files, retrieved from 23 different concert songs from YouTube, was used as the database of the audio fingerprinting algorithm, which corresponded to an average of 8.6 different recordings per concert song (*i.e.* event). This database generated 3098 matches, which were used to train, validate, and test our models. From these, there were 1071 true matches (class 1) and 2027 false matches (class 0) from which 2021 were repetition matches and 6 were wrong matches. Note that we balanced the training set every time a new model was trained (*i.e.* the number of samples of class 0 was equal to the number of samples of class 1).

### 4.3. Model Estimation

We used three different methods to solve this classification problem: logistic regression, k-nearest neighbours (kNN), and support vector machines (SVM). The purpose of using different classifiers is to have a broader way of comparison on how the different features used influence the outcome of the overall predictions of the different methods.

Apart from trying different feature vectors, we also varied the classifiers parameters. For logistic regression, we doubled the value of the regularisation parameter $c$ during 20 iterations (with its initial value being set to 1.0). We tried all odd numbers between 1 and 39 for the number of neighbours $k$ in the kNN classifier. Regarding the SVM classifier, we used the RBF kernel and the optimal values for $c$ and $\gamma$ were obtained by executing an exhaustive search over all possible combinations of a subset of possible values for each parameter. For this we followed the methodology of using exponentially growing sequences [4]. More specifically varying $c$ to the following values $2^{-5}, 2^{-3}, \ldots, 2^{15}, 2^{17}$ and $\gamma$ to $2^{-15}, 2^{-13}, \ldots, 2^{3}, 2^{5}$. This searching process is often described as Grid-search, and it returns the best value of each parameter of a given model (*i.e.* the hypothesis that achieves the highest accuracy).

We used double cross-validation to retrieve the model with lowest validation error for each classifier (varying the parameters as explained above): we start by performing **leave-one-song-out** cross-validation, in which every song in the training set except one are used to train the model with a $k$-fold cross-validation, with $k = 10$, whilst the left-out song is used to test the model; this process is then repeated until all songs have been left-out and repeated in every combination of possible parameters assigned for each classifier. The training and validation error of each model is the average of the error occurred in all the leave-one-song-out iterations, with the accuracy of the model being tested on the overall predictions of all left-out songs' samples. Following these steps for all designated ranges of possible values for the different classifiers' parameters, we assign the model with lowest validation error in the 10-fold validation for each classifier as the most suitable model.

## 4.4. Prediction Results

The accuracy results for each classifier is shown in figure 1. The SVM showed better results across the different feature combinations (98.23%, 97.22%, 96.12%, and 97.68%, respectively) but was closely followed by the other classifiers with the exception of logistic regression with ($\#ML_{o_{qi}}, \#L_{s_q}, \#L_{s_i}$), that achieved a considerably lower accuracy (82.07%).

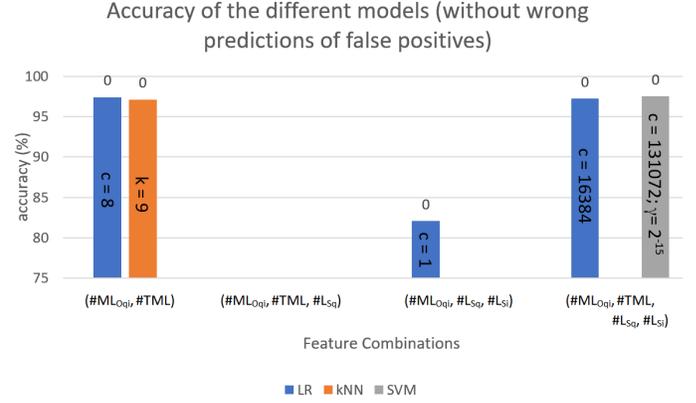

Fig. 2. The models that are missing in the grah incorrectly classified at least one wrong match. New models with different parameter values were found for both kNN and SVM whilst respecting this condition.

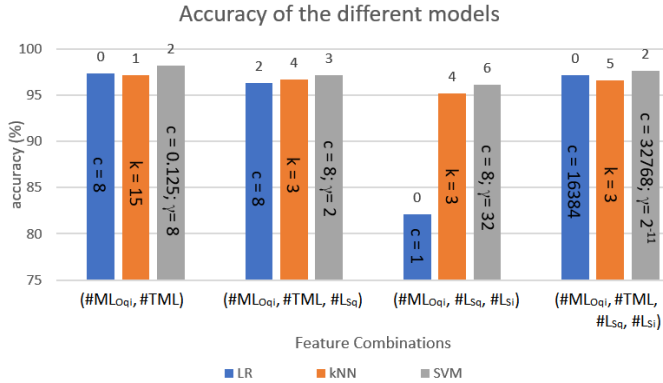

Fig. 1. Accuracy of the best models (*i.e.* with lowest validation error) of each classifier for the different combination of features. The parameter values are described inside each bar. The numbers placed on top of each bar represent the number of false positives for each model.

Despite their high accuracy, models that incorrectly classify wrong matches (false positives) have songs of different events assigned to the same cluster, leading ultimately to the merge of clusters of different events. Therefore, instead of simply choosing the model with lowest validation error for each classifier, we can discard all models that wrongly classified the wrong matches and choose the lower validation error model of the remaining. Figure 2 shows the updated classifiers results adding this constraint.

Even though the models' accuracy slightly decreased, we managed to find new models for kNN and SVM that satisfy our condition of classifying wrong matches correctly (that is, to class 0), whilst maintaining a high accuracy (97.12% and 97.49%, respectively). The logistic regression models already presented in figure 1 remained intact since they had no incorrect classifications of wrong matches, except when using the feature combination ($\#ML_{o_{qi}}, \#TML, \#L_{s_q}$), with their accuracy of 97.40% for the first presented feature combination, and 97.21% for the latter.

In sum, there is a slight advantage of considering only the models with no incorrect classification of wrong matches since their filtering is crucial in the proposed solution. We managed to achieve high accuracy results with each of the three classifiers. Using logistic regression and kNN with ($\#ML_{o_{qi}}, \#TML$), that is, the matching landmarks in the right offset and the total number of matching landmarks in all detected offsets, as well as using logistic regression and SVM with the 4 feature-combination, would represent practically viable options for the presented filtering approach.

## 5. LEARNING EXTENSION

The training set can be further expanded by the analysis of the information retrieved from the audio fingerprint algorithm combined with our model predictions. This extension can occur in two stages: during the audio clustering phase (section 2.2), and by the analysis of the matches between the cut samples when performing the audio quality inference inside each segment (section 3.3).

During the audio clustering phase, all repetition matches (repetitions of a given matched song in another's matching list) can be added to the training set: the feature vectors of the repetition matches are assigned to class 0. This is supported by the assumption that since only one offset is possible between two songs, the correct offset is the one that generated more matching landmarks, whilst the others are discarded.

The quality inference stage can serve as a confirmation for some of the samples that were predicted as true matches after the filtering method. Since all matches classified as wrong matches are filtered in the Audio Clustering phase (section 2.2), either by the discarding of the repetitions or by false matches classifications, all the samples of the songs present in the Audio Segmentation phase (section 3) were therefore predicted as true matches by our model (*i.e.* assigned to class 1). Hence, after cutting each song according to the different segments in which it appears, and by matching all cut songs with one another inside a segment to infer the quality, all matches should be assigned to offset 0.0 seconds since all the cut songs are meant to be synchronised in time.

Let us define the function *offset*($s, S$) as returning the set of offset values of the different matches between song $s$ and each song in set $S$. Function *count*$(A, v)$ retrieves the number of occurrences of the real number $v$ in set $A$, and $TM(m)$ assigns sample match $m$ to class 1 in the training set. Then, we can define the following expression for every cluster $c$:

$$\forall t \in T_c \ \forall s \in S_t \ count(\mathit{offset}(s, S_t \backslash \{s\}), 0) = \|S_t \backslash \{s\}\|$$
$$\Rightarrow \forall m \in M_c : TM(m)$$

where $t$ is a segment in $T_c$, which in turn is the set of (time) segments of cluster $c$, and $S_t$ is the set of songs in segment $t$. $M_c$ is the set of all matches (*i.e.* samples) in that cluster.

To sum up, one can then assume that, if for each cut song inside each cluster's segments there is a match to each of the other cut songs with offset of 0 seconds, then all samples that previously contributed to the formation of that given cluster are considered true matches and added to the training set with their class assigned to 1.

## 6. CONCLUSION

In this work we propose different methods that manipulate and correlate different user-generated recordings in a possibly large dataset of audio files, contributing ultimately for a better comprehension of the data. The basis of all presented work relied upon the direct analysis of the information retrieved by the matches of the different audio files from the audio fingerprinting algorithm.

Although using audio fingerprinting to organise different audio files with common audio excerpts was initially proposed by Kennedy and Naaman [5] and further extended in our previous work [7], here we introduced a novel filtering approach by using machine learning techniques and achieving optimal filtering results (*i.e.* successfully filtering all wrong matches) whilst also achieving high prediction accuracy in our considerable large test setup (*e.g.* $97.49\%$ using SVM and 4 features). Moreover, we introduce the possibility of extending our learning by increasing the training set in different possible stages, more concretely in the Audio Organisation and Audio Segmentation phases, by the detection of repetitions and by the analysis of the previous predictions.

We additionally proposed Audio Segmentation inside each cluster/event which provides valuable insight on how the different event's audio files are correlated in terms of time. This can be extremely useful since it provides the knowledge of which audio files are available at a given moment of time. Moreover, using a previously proposed audio inference approach [7] with parameter adaptations in the audio fingerprinting algorithm, we also represent how the different audio files relate in terms of their relative audio quality inside each segment of a given cluster.

## Acknowledgments

This work was partially funded by the H2020 ICT project COGNITUS with grant agreement No 687605 and by the Portuguese Foundation for Science and Technology under project NOVA-LINCS PEest/UID/CEC/04516/2013.